\documentclass[
    ,final            % use final for the camera ready runs
  ]
  {aipproc}

\layoutstyle{8x11double}

\begin{document}

\title{Metastable supersymmetry breaking vacua \\ from 
conformal dynamics\footnote{This talk is based on the paper~\cite{CFSB} collaborated with T. Kobayashi and H. Abe. Talk given by Y.O. at SUSY08 (Seoul, Korea, June 16-21, 2008)
} }


\author{Yuji~Omura\footnote{E-mail address: omura@scphys.kyoto-u.ac.jp
}}{
  address={Department of Physics, Kyoto University, 
Kyoto 606-8502, Japan}
}

\begin{abstract}
We study the scenario that conformal dynamics leads to 
metastable supersymmetry breaking vacua.
At a high energy scale, the superpotential is not R-symmetric, 
and has a supersymmetric minimum.
However, conformal dynamics suppresses several operators 
along renormalization group flow toward the infrared fixed point.
Then we can find an approximately R-symmetric superpotential, 
which has a metastable supersymmetry breaking vacuum, 
and the supersymmetric vacuum moves far away from 
the metastable supersymmetry breaking vacuum.
We show a 4D simple model.
Furthermore, we can construct 5D models with 
the same behavior, because of the AdS/CFT dual. 
\end{abstract}

\maketitle

%%%%%%%%%%%%%%%%%%%%%%%%%%%%%%%%%%%%%%%%%%%%
%% MAINMATTER
%%%%%%%%%%%%%%%%%%%%%%%%%%%%%%%%%%%%%%%%%%%%

\section{Introduction}
Conformal dynamics provides  several interesting 
aspects in supersymmetric models as well as 
non-supersymmetric models, because 
conformal dynamics exponentially suppresses or enhances 
certain operators.
One interesting aspect is that conformal dynamics can suppress flavor-dependent 
contributions to soft SUSY breaking terms which lead to flavor changing neutral current processes constrained strongly by current experiments.
Then, flavor-blind contributions such as 
anomaly mediation \cite{Randall:1998uk} would become dominant\cite{Luty:2001jh,Dine:2004dv,Sundrum:2004un,Ibe:2005pj,
Schmaltz:2006qs,Murayama:2007ge}.
Another interesting aspect is that conformal dynamics 
can generate hierarchical structure of Yukawa couplings 
for quarks and leptons \cite{Nelson:2000sn,Kobayashi:2001kz}.

Here we study a new application of conformal dynamics 
for supersymmetric models, that is, 
realization of metastable SUSY breaking vacua 
by conformal dynamics.
Its idea is as follows.
The Nelson-Seiberg argument \cite{Nelson:1993nf} implies 
that 
generic superpotential has a SUSY minimum, 
but R-symmetric superpotential has no SUSY minimum, 
that is, SUSY is broken in such a model.
Thus, if we add explicit R-symmetry breaking terms 
in R-symmetric superpotential, 
a SUSY minimum would appear.
However, when such R-symmetry breaking terms are 
tiny, the previous SUSY breaking minimum would 
survive and a newly appeared SUSY preserving minimum 
would be far away from the SUSY breaking point 
in the field space.
That is the metastable SUSY breaking vacuum 
\cite{Intriligator:2007py,Shih:2007av,Abe:2007ax}.
We try to realize such a metastable SUSY 
breaking vacuum by conformal dynamics.
We start with a superpotential without R-symmetry.
However, we assume the conformal dynamics.
Because of that, certain couplings are exponentially 
suppressed.
Then, we could realize an R-symmetric superpotential 
or an approximately R-symmetric superpotential 
with tiny R-symmetry breaking terms.
It would lead to a stable or metastable SUSY breaking vacuum.
We study this scenario by using a simple model.
Also, we study 5D models, which have the same behavior.

\section{4D conformal model}

Our model is the $SU(N)$ gauge theory with $N_f$ flavors 
of chiral matter fields $\phi_i$ and $\tilde \phi_i$, which are 
fundamental and anti-fundamental representations of 
$SU(N)$.
The flavor number satisfies $N_f \geq \frac32 N$, 
and that corresponds to the conformal window 
\cite{Seiberg:1994pq,Intriligator:1995au},  
that is, this theory has an IR fixed point \cite{Banks:1981nn}.
The NSVZ beta-function of physical gauge coupling 
$\alpha=g^2/8\pi^2$ is  
\begin{equation}
\beta^{\rm NSVZ}_\alpha= - \frac{\alpha^2}{1-N\alpha} (3N-N_f+N_f \gamma_\phi),
\end{equation}
where $\gamma_\phi$ is the anomalous dimension of 
$\phi_i$ and $\tilde \phi_i$ \cite{Novikov:1983uc, ArkaniHamed:1997ut}.
Since the IR fixed point corresponds to $\beta^{\rm NSVZ}_\alpha=0$, 
around that point the matter fields $\phi_i$ and $\tilde \phi_i$ have 
anomalous dimensions $\gamma_\phi= -(3N-N_f)/N_f$, which 
are negative.

In addition to the fields $\phi_i$ and $\tilde \phi_i$, 
we introduce singlet fields $\Phi_{ij}$ for $i,j=1,\cdots,N_f$.
The gauge invariance allows the following superpotential 
at the renormalizable level,
\begin{equation}
W=h\phi_i\Phi_{ij}\tilde \phi_j + f {\rm Tr}_{ij }\Phi_{ij} + 
\frac{m}{2} {\rm Tr}_{ik} \Phi_{ij}\Phi_{jk} + \frac{\lambda}{3} 
{\rm Tr}_{i\ell}\Phi_{ij}\Phi_{jk}\Phi_{k\ell}.
\label{W-1}
\end{equation}
Here we have preserved the $SU(N_f)$ flavor symmetry.
Even if the $SU(N_f)$ flavor symmetry is broken, e.g 
by replacing $f {\rm Tr}_{ij }\Phi_{ij}$ by $f_{ij}\Phi_{ij}$, 
the following discussions would be valid.
For simplicity, we assume that all of couplings, 
$h$, $f$, $m$, $\lambda$, are real, although the following 
discussions are available for the model with 
complex parameters, $h$, $f$, $m$ and $\lambda$.
We assume that the mass terms of $\phi_i$ and $\tilde \phi_j$ vanish.

If $m =\lambda =0$, the above superpotential 
corresponds to the superpotential of the 
Intriligator-Seiberg-Shih (ISS) model 
\cite{Intriligator:2006dd}.
We consider that our theory is an effective theory with 
the cutoff $\Lambda$.
We assume that dimensionless parameters $h$ and $\lambda$ 
are of $O(1)$ and dimensionful parameters 
$f$ and $m$ satisfy $f \approx m^2$ and $m \ll \Lambda$.
We denote physical couplings as 
$\hat h=(Z_\phi Z_{\tilde \phi} Z_\Phi)^{-1/2}h$, 
$\hat f_{ij} = (Z_\Phi)^{-1/2} f_{ij}$, 
$\hat m = (Z_\Phi)^{-1} m$ and $\hat \lambda = (Z_\Phi)^{-3/2} \lambda$, 
where $Z_\phi, Z_{\tilde \phi}, Z_\Phi$ are wavefunction renormalization 
constants for $\phi,\tilde \phi, \Phi$, respectively.

First, we study the behavior of this model around 
the energy scale $\Lambda$.
The F-flat conditions are obtained as 
\begin{eqnarray}
\partial_{\Phi_{ij}}W &=& h\phi_i \tilde \phi_j + f\delta_{ij} 
+ m \Phi_{ij} + \lambda \Phi_{jk}\Phi_{ki}
=0, \\
\partial_{\phi_i} W &=& h \Phi_{ij}\tilde \phi_j =0, \\
\partial_{\tilde \phi_j} W &=& h \phi_i \Phi_{ij} =0.
\end{eqnarray}
These equations have a supersymmetric solution for generic values of 
parameters, $h, m, \lambda$.
We decompose $\phi, \tilde \phi$ 
and $\Phi$ as 
\begin{equation}
\Phi = \left(
\begin{array}{cc}
Y & Z^T \\
\tilde Z & X
\end{array}
\right), 
\qquad 
\phi = \left(
\begin{array}{c}
\chi \\
\rho
\end{array}
\right),
\qquad 
\tilde \phi^T = \left(
\begin{array}{c}
\tilde \chi \\
\tilde \rho
\end{array}
\right),
\end{equation}
where $Y$, $\chi$ and $\tilde \chi$ are $N\times N$ matrices, 
$X$ is an $(N_F-N)\times (N_F-N)$ matrix, 
$Z$, $\tilde Z$, $\rho$ and $\tilde \rho$ are 
$(N_F-N)\times N$ matrices. Let us consider the slice with $Z=\tilde{Z}=\rho=0$,
and we find a supersymmetric solution,
 \begin{equation}
x_s=\frac{-m \pm \sqrt{m^2-4f\lambda}}{2\lambda},
\label{sol-x}
\end{equation}
and
\begin{equation}
f\delta_{ij} + h \chi_i \tilde \chi_j = 0, \qquad  Y_{ij}=0,
\label{sol-Y}
\end{equation}
where $x_s$ is defined as $X_{ij}=x_s\delta_{ij}$. In addition, the D-flat conditions correspond to 
$|\chi_i| = |\tilde \chi_i|$.

Now let us study the behavior around the IR region.
We assume that the gauge coupling is around the IR fixed point, 
i.e. $\beta_\alpha \approx 0$, and that $\phi_i$ and $\tilde \phi_i$ 
have negative anomalous dimensions $\gamma_\phi$.
In addition, we assume that the physical Yukawa coupling $\hat h$ 
is driven toward IR fixed points.
The beta-function of $\hat h$ is obtained as 
\begin{equation}
\beta_{\hat h} = \hat h (\gamma_\phi +\gamma_{\tilde \phi} + \gamma_\Phi).
\end{equation}
The condition of the fixed point leads to 
$2 \gamma_\phi + \gamma_\Phi =0$.
Since $\gamma_\phi < 0$, we obtain a positive anomalous dimension 
for $\Phi_{ij}$.
Then, physical couplings, $\hat f$, $\hat m$ and 
$\hat \lambda$, are suppressed exponentially 
toward the IR direction as 
\begin{eqnarray}
\hat f(\mu) &=& \left( \frac{\mu}{\Lambda} \right)^{\gamma_\Phi}
\hat f(\Lambda), \qquad   
\hat m(\mu) = \left( \frac{\mu}{\Lambda} \right)^{2 \gamma_\Phi}
\hat m(\Lambda), \nonumber \\   
\hat \lambda(\mu) &=& \left( \frac{\mu}{\Lambda} \right)^{3\gamma_\Phi}
\hat \lambda(\Lambda).
\end{eqnarray}
Thus, the mass parameter $\hat m$ and 3-point coupling $\hat \lambda$ 
are suppressed faster than $\hat f$.
If we neglect $\hat m$ and $\hat \lambda$ but not 
$\hat f$, the above superpotential becomes 
the superpotential of the ISS model, and 
there is a SUSY breaking minimum around $\Phi_{ij}=0$ because of 
the rank condition.
We consider the overall direction, $X_{ij}=x\delta_{ij}$, and we use the 
canonically normalized basis, $\hat x$. We add the mass term $m_x^2|\hat x|^2$ 
in the one-loop effective potential
and analyze the potential, $V= V_{\rm SUSY} + m_x^2|\hat x|^2$ around $\hat x=0$.

Eventually, at a high energy scale corresponding to $Z_\Phi=O(1)$, 
we have $|\hat f|, |\hat m|^2 \gg m^2_x$, because 
$m_x^2$ is smaller than $\hat f$ by a loop factor.
The potential and the stationary condition are 
controlled by $|\hat f|, |\hat m|^2$, $\hat \lambda$, 
but not $m_x$.
Thus, there is no (SUSY breaking) minimum around $x=0$, 
but we have a supersymmetric minimum (\ref{sol-x}). 
However, toward the IR direction, 
$\hat m^2$ becomes suppressed faster than $m_x^2$.
Then, the couplings $\hat f$ and $m_x^2$ are important 
in the potential so that we find a metastable SUSY breaking vacuum
around $\hat x=0$, 
\begin{equation}
\hat x_{sb} \approx - \frac{\hat f \hat m}{m_x^2}.
\end{equation}

Both breaking scales of the $SU(N)$ gauge symmetry and supersymmetry 
at the metastable SUSY breaking point $\hat x =0$ are 
determined by $O(\hat f(\mu))$.
Thus, such an energy scale is estimated as 
$\mu_{IR}^2 \sim \hat f(\mu_{IR})$, i.e. 
\begin{equation}
\mu_{IR} \sim \left( \frac{\hat f(\Lambda)}{\Lambda^{\gamma_\Phi}} 
\right)^{1/(2-\gamma_\Phi)},
\end{equation}
and at this energy scale conformal renormalization group flow 
is terminated.

So far, we have assumed that the mass term of 
$\phi_i$ and $\tilde \phi_i$, $m_\phi \phi_i \tilde \phi_i$ vanishes.
Here, we comment on the case with such terms. 
The physical mass $\hat m_\phi$ becomes enhanced as 
\begin{equation}
\hat m_\phi(\mu) = \left( \frac{\mu}{\Lambda} \right)^{2 \gamma_\phi} 
 \hat m_\phi(\Lambda),
\end{equation}
because of the negative anomalous dimension $\gamma_\phi$.
At $\mu \sim \hat m_\phi(\mu)$, the matter fields $\phi_i$ $\tilde \phi_i$
decouple and this theory removes away from the conformal window.
Thus, if $\hat m_\phi(\mu) > \mu_{IR}$, the conformal renormalization 
group flow is terminated at $\mu_D \sim  \hat m_\phi(\mu_D)= 
(\mu_D/\Lambda)^{2\gamma_\phi}\hat m_\phi(\Lambda)$.

We have studied the scenario that 
conformal dynamics leads to metastable SUSY breaking 
vacua.
As an illustrating example of our idea, 
we have used the simple model.
Our scenario could be realized by other models.

\section{5D model}

There would be an AdS dual to our conformal scenario.
Indeed, we can construct simply various models 
within the framework of 5D orbifold theory.
Renormalization group flows in the 4D theory correspond to 
exponential profiles of zero modes like $e^{-c_i Ry}$, where 
$R$ is the radius of the fifth dimension,\footnote{
We assume that the radion is stabilized.} 
$y$ is the coordinate for the extra dimension, i.e. $y=[0,\pi]$ and 
$c_i$ is a constant.
The parameter $c_i$ corresponds to anomalous dimension in the 4D theory,
and each field would have a different constant $c_i$.
In 4D theory, values of anomalous dimensions are constrained 
by concrete 4D dynamics.
However, constants $c_i$ do not have such strong constraints, 
although they would correspond to some charges.
Hence, 5D models would have a rich structure 
and one could make model building rather simply.
In Ref.~\cite{CFSB}, we show a simple 5D model compactified on $S^1/Z_2$.

\section{Conclusion and discussion}

We have studied the scenario that conformal dynamics leads to 
approximately R-symmetric superpotential with a metastable 
SUSY breaking vacuum.
We have shown a simple model to realize our scenario. 
At a high energy scale, there would be 
only SUSY minimum and at low energy metastable 
SUSY breaking vacuum would appear.

We can make 5D models with the same behavior.
Since in our 4D scenario, metastable SUSY breaking vacua 
are realized by conformal dynamics, 
such a SUSY breaking source would be sequestered from the 
visible sector by conformal dynamics.
%Similarly, in 5D models a SUSY breaking source would be 
%sequestered from the visible sector by geometrical sequestering.

\begin{theacknowledgments}
The author would like to thank H. Abe and T. Kobayashi for fruitful collaborations.
The work of Y.O. is supported by the Japan 
society of promotion of science (No. 20$\cdot$324). 
\end{theacknowledgments}

\end{document}